\documentclass[pra,twocolumn,preprintnumbers,amsmath,amssymb,floatfix,showpacs]{revtex4}
\usepackage{hyperref}
\usepackage{amsmath}
\usepackage[dvips]{epsfig}

\begin{document}

\title{Path-Integral Ground-State and Superfluid Hydrodynamics \\
of a Bosonic Gas of Hard Spheres}
\author{Maurizio Rossi and Luca Salasnich} 
\affiliation{Dipartimento di Fisica e Astronomia 
"Galileo Galilei" and CNISM, Universit\`a di Padova, 
Via Marzolo 8, 35122 Padova, Italy} 

\begin{abstract} 
We study a bosonic gas of hard spheres by using of the exact zero-temperature 
Path-Integral Ground-State (PIGS) Monte Carlo method and the equations of 
superfluid hydrodynamics. 
The PIGS method is implemented to calculate for the bulk system the energy 
per particle and the condensate fraction through a large range of the gas 
parameter $na^3$ (with $n$ the number density and $a$ the s--wave scattering 
length), going from the dilute gas into the solid phase. 
The Maxwell construction is then adopted to determine the freezing at 
$na^3=0.278\pm 0.001$ and the melting at $na^3=0.286\pm 0.001$. 
In the liquid phase, where the condensate fraction is finite, the equations 
of superfluid hydrodynamics, based on the PIGS equation of state, are used to
find other relevant quantities as a function of the gas parameter: the 
chemical potential, the pressure and the sound velocity. 
In addition, within the Feynman's approximation, from the PIGS static structure 
factor we determine the full excitation spectrum, which displays a maxon-roton
behavior when the gas parameter is close to the freezing value. 
Finally, the equations of superfluid hydrodynamics with the PIGS equation of 
state are solved for bosonic system under axially--symmetric harmonic 
confinement obtaining its collective breathing modes. 
\end{abstract} 

\date{\today}

\pacs{02.70.Ss, 03.75.Hh, 03.75.Kk}

\maketitle

\section{Introduction}

In this paper we analyze a system of identical interacting bosons by using the
hard sphere (HS) model \cite{hans2}, which is a useful reference system for 
classical and quantum many-body theories both for weak and strong interactions 
because it depends only on one interaction parameter: the sphere diameter $a$ 
\cite{hans2,book1,book2}. 
The quantum HS model has led to the understanding of several general features 
of helium in its condensed phases \cite{book1,book2}, serving as a reference or
a starting point for studies with more accurate potentials \cite{kalo}. 
In addition the quantum HS model provides the standard benchmark for 
mean--field approaches \cite{boro2} such as, for example, Gross-Pitaevskii 
equation or Hartree--Fock--Bogoliubov approximation \cite{kim}. 
A large number of approaches have been put forward to deal with quantum HS and, 
among them, Monte Carlo method based on Feynman`s path integrals come out as a 
most powerful tool \cite{sese2}. 
Path integral Monte Carlo (PIMC) studies of quantum HS systems at finite 
temperature cover almost the whole relevant gas parameter range 
\cite{ches,sese,sese2,sese3}. 
However, at zero temperature, there are studies that cover (with different 
techniques) only portions of the $na^3$ range and are mainly devoted to the 
investigation of different properties such as the universal behavior in the 
dilute limit \cite{boro2} or the gas--solid transition \cite{kalo}.

Here we calculate the equation of state of the bulk quantum HS system of 
identical bosons from very low gas parameter value up to high density solid with 
path integral ground state (PIGS) Monte Carlo \cite{pigs} method, which provides 
exact expectation values on the ground state. 
Our exact PIGS results for the equation of state are then used to derive other 
relevant properties by means of the equations of superfluid hydrodynamics 
\cite{book1,book2}. 
Actual experiments on Bosonic atomic gases reach so low temperatures that the
effects of thermal fluctuations are largely negligible, making a zero temperature
approach well justified\cite{book1,book2}.
The paper is organized in the following way. 
The basic features of PIGS method are reported in Section II. 
Numerical results on the ground-state energy and condensate fraction are shown 
and discussed in Section III, where we compare our data with previous Monte Carlo
calculations and other theoretical approaches. 
In Section IV we introduce the zero-temperature hydrodynamic equations of 
superfluids \cite{book1,book2} and we use them (with the PIGS equation of state) 
to find other relevant quantities as a function of the gas parameter: the chemical
potential, the pressure and the sound velocity. 
We find that our sound velocity, which gives the low-momentum linear slope of the 
excitation spectrum, is in excellent agreement with the numerical results obtained
with the help of the PIGS static response function. 
Moreover, within the Feynman's approximation, we determine the full spectrum of 
elementary excitations, which displays a maxon-roton behavior when the gas 
parameter is close to the freezing value. 
In section V we consider the inclusion of an anisotropic but axially-symmetric
trapping harmonic potential. 
The collective modes of the confined Bose gas are then easily calculated using 
again the equations of superfluid hydrodynamics with the PIGS equation of state, 
which is locally approximated with a polytropic equation of state \cite{nick}. 
The paper is concluded by Section VI.

\section{PIGS method}

The aim of PIGS is to improve a variationally optimized trial wave function 
$\psi_t$ by constructing, in the Hilbert space of the system, a path which 
connects the starting $\psi_t$ with the exact lowest energy wave function of 
the system, $\psi_0$, constrained by the choice of the number of particles $N$, the 
geometry of the simulation box, the boundary conditions and the density $n$, 
provided that $\langle\psi_t|\psi_0\rangle\neq 0$. 
The correct correlations among the particles arise during this path through the 
action of the imaginary time evolution operator $\hat G=e^{-\tau\hat H}$, where 
$\hat H$ is the Hamiltonian operator.
In principle, $\psi_0$ is reached in the limit of infinite imaginary time, but a 
very accurate representation for $\psi_0$ is given by 
$\psi_{\tau}=e^{-\tau\hat H}\psi_t$, if $\tau$ is large enough (but finite).

The wave function $\psi_{\tau}$ can be analytically written by discretizing the path
in small imaginary time steps.
This discretization is necessary since the aviable approximations for $\hat G$ 
became more accurate as the imaginary time step goes smaller \cite{cepe}.
Here we have used the Cao--Berne approximation \cite{caob}, which is one of the most 
efficient propagators (i.e. allows for larger values of imaginary time step) for HS 
\cite{sese}.
Because because of this discretization of the imaginary time path, the quantum system
is mapped into a system of specially interacting classical open polymers \cite{pigs}.
Each open polymer represents the full imaginary time path of a quantum particle that 
is sampled by means of the Metropolis algorithm. 
Thus, the entire imaginary time evolution of the system is sampled at each Monte Carlo
step \cite{pata}.

An appealing feature of the PIGS method is that, in $\psi_{\tau}$, the variational 
ansatz acts only as a starting point, while the full path is governed by $\hat G$, 
which depends only on the Hamiltonian $\hat H$.
Thus the PIGS method turns out to be unbiased by the choice of the trial wave function 
\cite{pata} and then the only input is $\hat H$.
In the coordinate representation, the Hamiltonian of the quantum HS system is
\begin{equation}
 \label{H}
 H = -\frac{\hbar^2}{2m}\sum_{i=1}^N\nabla^2_i + \sum_{<i,j>}V(r_{ij})
\end{equation}
where $r_{ij}=|\vec r_i-\vec r_j|$ and
\begin{equation}
 \label{HS}
 V(r)=\left\{
             \begin{array}{ll}
             +\infty & {\rm for }\quad r<a\\
             0       & {\rm otherwise} \; .
             \end{array}
      \right.
\end{equation}
The Hamiltonian (\ref{H}) can be reduced in a useful adimensional form by giving the 
energies in unit of $\frac{\hbar^2}{2ma^2}$ and the lengths in unit of $a$, which 
represents also the s--wave scattering length.
We make use of these reduced units throughout the paper.

The trial wave function $\psi_t$ does not really need to be fully variational optimized: 
in fact, for a large enough value of $\tau$, PIGS results turn out to be independent on 
$\psi_t$, in both the phases \cite{pata,vita}.
The sole role of $\psi_t$ is to determine the length of the path in imaginary time 
\cite{pata} to converge on $\psi_0$: better is $\psi_t$, faster is the convergence.
Here, as $\psi_t$, we have employed a Jastrow wave function, where the two-body 
correlations are given by the fist order expansion of the exact solution for the two-body
problem i.e.:
\begin{equation}
 \label{psit}
 \psi_t(R) = \prod_{<i,j>}\left(1-\frac{a}{r_{ij}}\right)
\end{equation}
where $R=\{\vec r_1,\dots,\vec r_N\}$ are the coordinates of the $N$ HS.

All the approximations involved in the PIGS method, i.e. the choice of the total imaginary
time $\tau$ and of the imaginary time step $\delta\tau$ (that fixes the quality of the 
approximation on $\hat G$), are so well controlled that the resulting systematic errors
can be reduced within the unavoidable Monte Carlo statistical error.
In this sense PIGS is an exact $T=0$ K method \cite{pigs,pata}.

In order to improve the ergodicity of the Monte Carlo sampling, we have implemented bosonic
permutations \cite{boni}, even if not required in principle since (\ref{psit}) has the 
correct Bose symmetry, and a canonical (i.e. with fixed $N$) version of the worm algorithm 
\cite{worm}, which has the ulterior advantage of giving access also to off-diagonal 
properties within the same simulation.

\section{Ground-state energy and condensate fraction}

We have studied with PIGS a system of $N=256$ HS in a cubic box with periodic boundary 
conditions in all the directions, for values of the gas parameter $na^3$ ranging from dilute
gas, namely $na^3 = 10^{-3}$, up to $na^3 = 0.5$, deep inside the solid phase.
By studying the convergence in $\tau$ and $\delta\tau$ of the energy per particle we have
fixed the values $\tau = 0.225$~$2ma^2/\hbar^2$ and $\delta\tau = 0.015$~$2ma^2/\hbar^2$ to 
be a very good compromise between accuracy and computational cost.
For some values of $na^3$, we have check the convergence of our results both by reducing the 
time step to $\delta\tau = 0.005$~$2ma^2/\hbar^2$ and by extending the total projection time
up to $\tau=0.245$~$2ma^2/\hbar^2$.
We have performed also simulations with $N=400$ and $N=500$ HS in order to verify the presence
of size effects, especially close to the gas-solid transition region.
We find that the energy per particle does not sensibly change within the error bars, inferring
that our results are no affected by a significant size effect.

\begin{figure}
 \epsfig{file=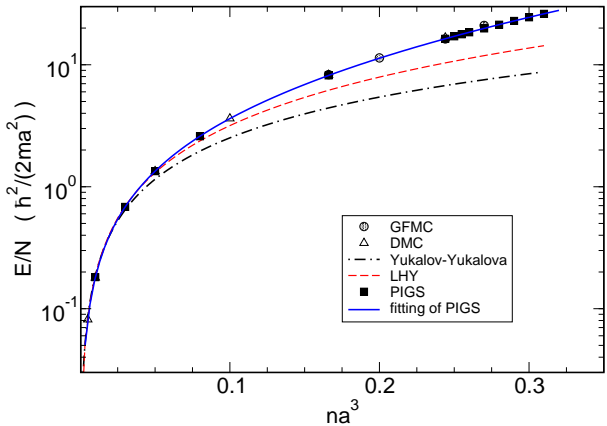,width=9 cm,clip=}
 \caption{(Colors online) Energy per particle $E/N$ in unit of $\hbar^2/2ma^2$ in 
          the gas phase as a function of the gas parameter $na^3$ computed with PIGS 
          (filled squares) compared with previous GFMC \cite{kalo} (shaded circles) 
          and DMC \cite{boro2} (open triangles) results. 
          Error bars are smaller than the used symbols.
          Dashed line: LHY perturbative approach \cite{lhy}; dot-dashed line: 
          Yukalov--Yukalova improved perturbative approach \cite{yuka}; 
          solid line: fit of PIGS data.}
 \label{liquid}
\end{figure}

Our results for the energy per particle $E/N$ as a function of the gas parameter are reported
in Fig.~\ref{liquid}. 
We find an excellent agreement with previous GFMC \cite{kalo} and DMC data \cite{boro2} in the
range of gas parameter values covered by the previous studies. 
We report also two mean field predictions for $E/N$: the perturbative correction to the 
Bogoliubov mean-field due to Lee, Huang and Yang (LHY) \cite{lhy} that turns out to be in a 
quite fair good agreement with Monte Carlo data up to $na^3\simeq5\times10^{-2}$ \cite{boro1}, 
and a more recent perturbative approach due to Yukalov and Yukalova \cite{yuka} that, however,
distances itself from Monte Carlo data for just lower values of $na^3$.
 
By adding successive powers at the mean-field prediction of Bogoliubov with the LHK perturbative
correction, we have fit our data with the expression \cite{boro1} 
\begin{equation}
 {E\over N} = {\hbar^2\over 2 m a^2} \, f_g(na^3) \; , 
\end{equation}
where 
\begin{equation}
 \label{liq_fit}
 \begin{split}
    f_g(x) =& 4\pi x\left(1+\frac{128}{15\sqrt{\pi}}\sqrt{x}\right) 
            + a_2x^2\log(x) + b_2x^2 \\ 
          & + a_{5/2}x^{5/2}\log(x) + b_{5/2}x^{5/2} \; .  
 \end{split}
\end{equation}
The best values for the parameters coming from the fit of the PIGS data are $a_2=145.5$, 
$b_2=842.8$, $a_{5/2}=422$ and $b_{5/2}=-492$.
The resulting curve of $f_g(x)$ is also reported as a solid line in Fig.~\ref{liquid}.

\begin{figure}
 \epsfig{file=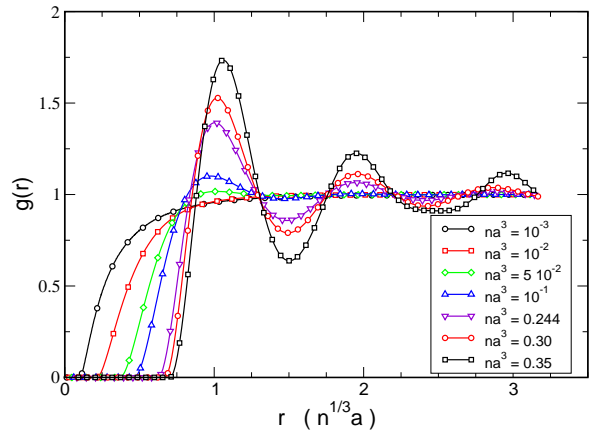,width=9 cm,clip=}
 \caption{(Colors online) Pair correlation function $g(r)$ for different values of
          the gas parameter $na^3$ conputed with PIGS.
          Error bars are smaller than the used symbols.}
 \label{gdir}
\end{figure}

\begin{figure}
 \epsfig{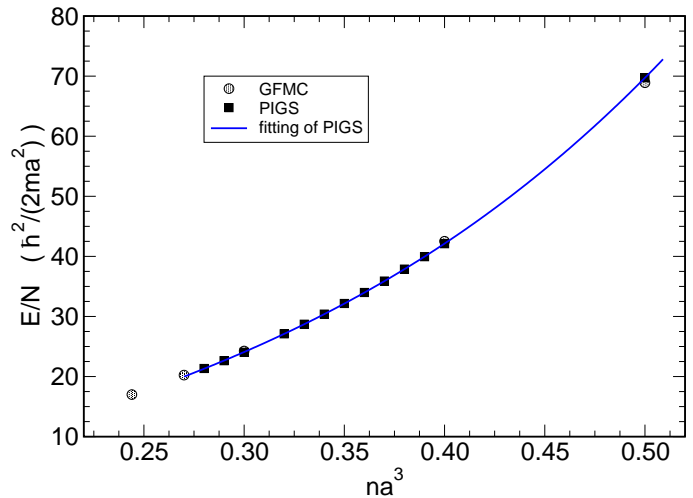}
 \caption{(Colors online) Energy per particle $E/N$ in unit of $\hbar^2/2ma^2$ in 
          the solid phase as a function of the gas parameter $na^3$ computed with 
          PIGS (filled squares) compared with previous GFMC \cite{kalo} (shaded 
          circles) results. 
          Error bars are smaller than the used symbols.
          Solid line: fit of PIGS data.}
 \label{solid}
\end{figure}

By increasing the gas parameter the system spontaneously breaks the translational invariance
due to the effect of the increased correlations among the particles, resulting in a solid phase, 
as inferred also from the characteristic oscillations in the pair correlation function
\begin{equation}
 \label{eq_gr}
 g(r) =  \frac{N(N-1)}{n^2}
         \frac{\int\prod_{j=3}^N d\vec r_j\:\left|\psi^*_0(\vec r,0,\vec r_3,\dots,\vec r_N)\right|^2}
              {\int\prod_{j=1}^N d\vec r_j\:\left|\psi^*_0(\vec r_1,\vec r_2,\dots,\vec r_N)\right|^2}
\end{equation}
reported in Fig.~\ref{gdir}.
The emerging crystal is the FCC, that is the lattice that best fits the cubic geometry of the 
simulation box.
Very recent PIMC simulations \cite{sese2} have shown, however, that the free energy difference 
between the two closed packed crystals, FCC and HCP, for the quantum HS, is vanishing small.
This is not surprising since the difference in these two lattices arises from the second shell 
of neighbors, and the HS potential is short ranged. 
In Fig.~\ref{solid} we report the resulting energy per particles $E/N$ as a function of the gas
parameter. 
Even in this case, we find a quite good agreement with older GFMC data \cite{kalo}.
We find that our results can be well fitted with a standard third-order polynomial \cite{hans}
\begin{equation} 
 {E\over N} = {\hbar^2\over 2 m a^2} \, f_s(na^3) \; , 
 \label{liq_fit0}
\end{equation}
where 
\begin{equation}
 \label{sol_fit}
 f_s(x) = E_0 + Ax + Bx^2 + Cx^3 \; , 
\end{equation}
and the best values for the fit parameters are $E_0=-9.33$, $A=132.6$, $B=-253.6$ and $C=609.1$. 
The resulting $f_s(x)$ is plotted as a solid line in Fig.~\ref{solid}.

By using the polynomial fit to the PIGS data (\ref{liq_fit}) and (\ref{sol_fit}) it is possible 
to locate the transition region between the gas and the solid phase via the standard Maxwell 
(double tangent) construction. 
We find that the coexistence region is bounded by $n_fa^3=0.264\pm0.003$ (freezing gas parameter)
and $n_ma^3=0.290\pm0.003$ (melting gas parameter).
These values are close, but not perfectly compatible, with the older GFMC results \cite{kalo} 
$n_fa^3=0.25\pm0.01$ and $n_ma^3=0.27\pm0.01$.
The shift to higher values for the bounding gas parameters can be due to a greater accuracy of 
the imaginary time propagator used here \cite{sese2}.
Another source of difference can be the strong dependence of such bounding values on the different 
used fitting formula, even if the energies $E/N$ obtained with the two exact Monte Carlo methods 
are very close (as one expects from exact techniques).

\begin{figure}
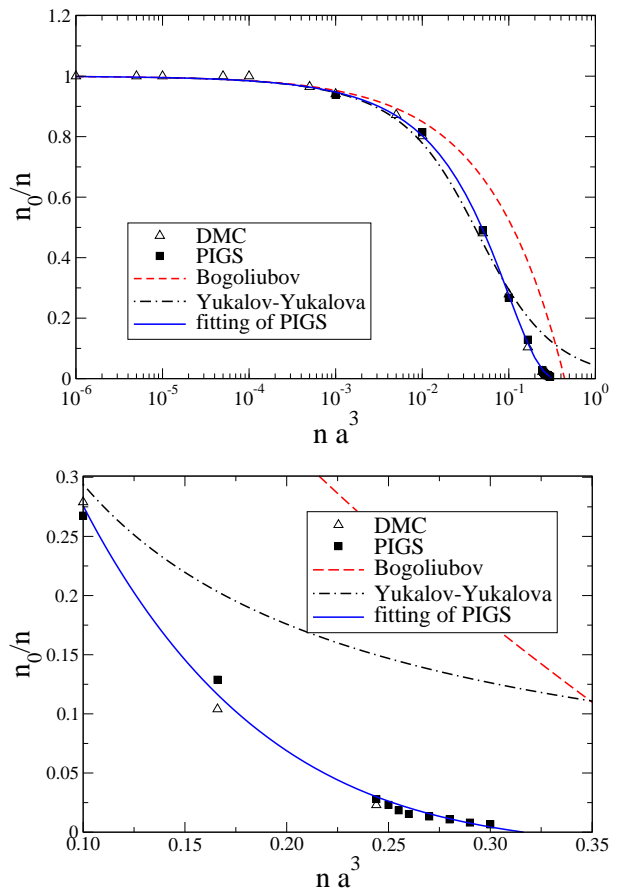

 \epsfig{file=fig4u.eps,width=8 cm,clip=}
 \vskip 0.3cm
 \epsfig{file=fig4l.eps,width=8 cm,clip=}
 \caption{(Colors online) Upper panel: condensate fraction $n_0/n$ as a function of 
          the gas parameter $na^3$, computed with PIGS (filled squares) compared 
          with the previous DMC results \cite{boro2} (open tringles). 
          Error bars are smaller than the used symbols. 
          Dashed line: Bogoliubov formula; 
          dot-dashed line: Yukalov-Yukalova improved perturbative approach; 
          solid line: fit of PIGS data.
          Lower panel: zoom of the upper panel showing DMC and PIGS data in the 
          region where the condensate fraction is going to zero and perturbative 
          methods fail.}
 \label{cond}
\end{figure}

The worm algorithm \cite{worm}, give direct access also to the one-body density matrix
\begin{equation}
 \label{rho1}
 \rho_1(\vec r,\vec r') = \int \prod_{j=2}^N d\vec r_j\: 
                          \psi^*_0(\vec r,\vec r_2,\dots,\vec r_N)
                          \psi_0(\vec r',\vec r_2,\dots,\vec r_N)
\end{equation}
that in a uniform system turns out to be function only of the difference $|\vec r - \vec r'|$.
$\rho_1$ is the Fourier Transform of the momentum distribution of the system, then 
a finite plateau in the large distance tail of $\rho_1$ means a Dirac's delta in 
the zero momentum state, i.e. a macroscopic occupation of a single particle quantum 
state that is the Bose--Einstein condensation.
The condensate fraction $n_0/n$ turns out to be equal to the limiting value of the 
tail of the one body density matrix.
We plot our results for $n_0/n$ in Fig.~\ref{cond}.
In the solid phase the condesate fraction turns out to be zero, in agreement with
what found in $^4$He systems \cite{worm,vita}.
In the gas phase, even for the condensate fraction we find a satisfactory agreement 
with previous DMC results \cite{boro2} in the $na^3$ range where they were aviable.
Our data confirm that the Bogoliubov prediction overestimates the condensate 
fraction for gas parameter larger than $na^3\simeq10^{-3}$ \cite{boro2}.
The improved perturbative approach of Ref.~\cite{yuka} gives a better prediction of 
$n_0/n$ starting to overestimate the condensate fraction for values of the gas 
parameter larger than $10^{-1}$, as shown in Fig.~\ref{cond}. 
To provide an analytical expression for the condensate fraction as a function of the 
gas parameter, we follow Ref.~\cite{boro1} and fit our data with the formula 
\begin{equation}
{n_0\over n} = \Xi(na^3) \; , 
\end{equation}
where 
\begin{equation} 
 \label{cond_fit}
 \Xi(x) = 1 - \frac{8}{3\sqrt{\pi}}\sqrt{x} 
            - c_1 x - c_{3/2} x^{3/2}
            - c_2 x^2 - c_{5/2}x^{5/2} \; .
\end{equation}
The best values for the fit parmeters are $c_1=5.49$, $c_{3/2}=-7.86$, $c_2=-9.52$ and 
$c_{5/2}=13.65$.

\section{Superfluid hydrodynamics and elementary excitations}

\begin{figure}
 \epsfig{file=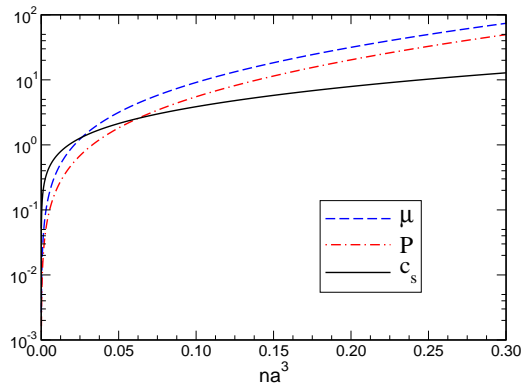,width=8.cm,clip=}
 \caption{(Colors online) Various physical quantities as a function of the gas 
          parameter $na^3$: the bulk chemical potential $\mu$ (in units of
          $\hbar^2/(2ma^2)$), the bulk pressure $P$ (in units of $\hbar^2/(2ma^2n)$), 
          and the sound velocity $c_s$ (in units of $(\hbar^2/(2m^2a^2))^{1/2}$).}
 \label{bulk}
\end{figure}

The advantage of a functional parametrization $f_g(x)$, Eq. (\ref{liq_fit}), of the 
ground-state energy $E$ of the bosonic gas is that it allows straightforward analytical
calculations of several physical properties \cite{nick}. 
For example, the bulk chemical potential $\mu$ is given 
\begin{equation}
 \mu = {\partial E\over \partial N} = {\hbar^2\over 2m a^2} \, 
 \left( f_g(x) + x f_g'(x) \right) \; , 
 \label{echem}
\end{equation}
as found by using Eqs. (\ref{liq_fit0}) and (\ref{liq_fit}) and taking into account that 
$x=na^3$ and ${\partial x}/{\partial n}=x/n$, while the bulk pressure $P$ reads 
\begin{equation}
 P = n^2 {\partial \over \partial n}\left( {E\over N} \right) 
   = {\hbar^2\over 2m a^2} \, n \, x \, f_g'(x) \; . 
\end{equation} 
Moreover, the collective dynamics of our bosonic gas of HS with local density 
$n({\bf r},t)$ and local velocity ${\bf v}({\bf r},t)$ can be described by the following 
zero-temperature hydrodynamic equations of superfluids \cite{book1,book2}
\begin{eqnarray}
  {\partial n\over \partial t} + {\boldsymbol \nabla} \cdot 
  \left( n \, {\bf v} \right) = 0 \; , 
  \label{dyn1}
 \\
  m {\partial {\bf v}\over \partial t} + {\boldsymbol \nabla} 
  \left[{1\over 2} m v^2 + \mu[n,a] \right] = {\bf 0} \; ,  
  \label{dyn2}
\end{eqnarray}
where $\mu[n,a]$ is the bulk chemical potential, given by Eq. (\ref{echem}). 
These equations describe a generic fluid at zero temperature which is 
inviscid (zero viscosity) and irrotational (${\bf v}\wedge {\bf v}={\bf 0}$) 
\cite{book1,book2}. The irrotationality implies that ${\bf v}={\boldsymbol \nabla} 
\theta$, where 
$\theta=\theta({\bf r},t)$ is a scalar field which must be an angle variable 
to get the quantization of the circulation of the velocity \cite{book1,book2}. 
Thus, from the knowledge of the bulk equation of state (\ref{echem}) one can study 
the collective superfluid dynamics of the system by solving Eqs. (\ref{dyn1}) and (\ref{dyn2}). 
In particular, we are interested on the propagation of sound waves in the superfluid. 
In this case, by taking into account a small $\delta n({r},t)$ variation of the local 
density with respect to the uniform value $n$ and linearizing the hydrodynamic equations
one finds the familiar wave equation 
\begin{equation}
 \left[ {\partial^2\over \partial t^2} - c_s^2 \, \nabla^2 \right] 
 \delta n({\bf r},t) = 0 \; , 
\end{equation}
where $c_s$ is the sound velocity, given by 
\begin{equation} 
 m c_s^2 = n \, {\partial \mu\over \partial n} = 
          {\hbar^2\over 2m a^2} \left( 2x f_g'(x) + x^2 f_g''(x) \right) \; . 
 \label{cs}
\end{equation}
It is well known that this wave equation admits monochromatic plane--wave solutions, 
where the frequency $\omega$ and the wave vector ${\bf k}$ are related by the phononic 
dispersion formula 
\begin{equation} 
 \hbar \omega(k) = c_s \, \hbar k \; ,  
 \label{linear}
\end{equation}
where $k=|{\bf k}|$ is the wavenumber. 
In Fig. \ref{bulk} we plot the bulk chemical potential $\mu$, the bulk pressure $P$ and 
the sound velocity $c_s$ as a function of the gas parameter $na^3$. 
All these physical quantities are calculated on the basis of the parametrization 
(\ref{liq_fit0}) and (\ref{liq_fit}) of the PIGS energy $E$. 

\begin{figure}
 \epsfig{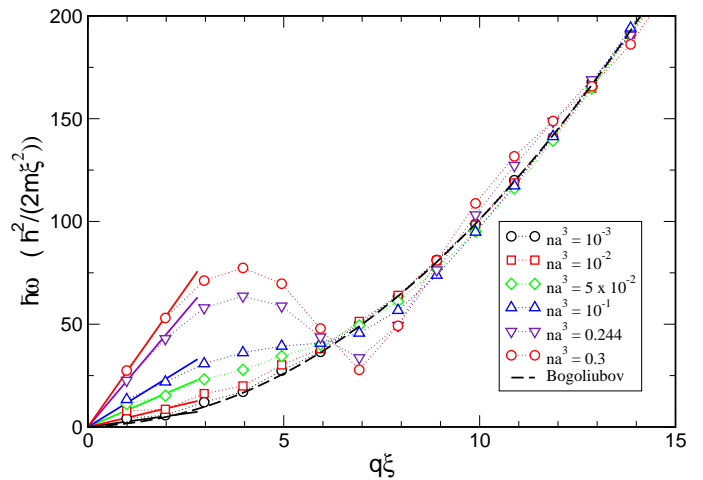}
 \caption{(Colors online) Excitation spectrum obtained with the Feynman's 
          approximation at different values of the gas parameter $na^3$.
          Dotted lines are given as guides to the eye.
          In order to make comparable results at different $na^3$ the spectra are 
          plotted as a function of $q\xi$ and in units of $\hbar^2/2m\xi^2$ where 
          $\xi=an^{1/3}$.
          In these units, the Bogoliubov approximation for the excitation spectrum 
          (dashed line) reads $\hbar\omega(q) = \sqrt{(q\xi)^4 + 2(q\xi)^2}$.
          The relative low wave vector phononic dispersion, Eq. (\ref{linear}) are
          reported as stright lines.}
 \label{spett}
\end{figure}

The zero-temperature equations of superfluid hydrodynamics (\ref{dyn1}) and (\ref{dyn2}), 
equipped by the constitutive equation of state (\ref{echem}) which is based on the 
parametrization (\ref{liq_fit0}) and (\ref{liq_fit}) of the PIGS energy, give reliable 
informations only on the low wavenumber branch (linear part) of the spectrum $\omega(k)$ 
of the elementary excitations. 
Unfortunately, the imaginary-time formulation of PIGS method prevent us from obtaining 
the exact dynamical properties of the system, such us the full excitation spectrum 
$\omega(k)$, directly from simulations. 
Some features of $\omega(k)$ can be obtained within the Feynman's approximation:
\begin{equation}
 \label{fey}
 \hbar\omega(k) = \frac{\hbar^2k^2}{2mS(k)}
\end{equation}
where 
\begin{equation}
 \label{static}
 S(k) = \frac{1}{N}\langle\sum_{j=1}^Ne^{-i\vec k\cdot\vec r_j}
        \sum_{l=1}^Ne^{i\vec k\cdot\vec r_l}\rangle 
\end{equation}
is the static structure factor that can be readily obtained during a PIGS simulation. 
Our results for the Feynman's excitation spectrum for HS at different values of the gas 
parameter are reported in Fig.~\ref{spett}. 
The Feynman's approximation is known to be accurate only at very low $na^3$, and to 
become only qualitative at higher values of the gas parameter. 
For example, in the case of superfluid $^4$He, where $na^3=0.244$, it overestimates the 
roton minimum by a factor of about two.
In the low wave vector limit we find that, in spite of the well known size effect on the 
static structure factor (\ref{static}) \cite{drae}, the Feynman's approximation turn out 
to be in a remarkable agreement with the phononic dispersion (\ref{linear}) with the 
values of the sound velocity $c_s$ given by Eq. (\ref{cs}) and reported in Fig.~\ref{bulk}. 
It is worthy to note that even the Feynman's approximation (\ref{fey}) for the excitation 
spectrum, as the energy per particle and the condensate fraction, deviates from the 
Bogoliubov approximation for $na^3\simeq 10^{-3}$.
Another remarkable feature is that, even within this simple approximation, the occurrence 
of a roton minimum at high density is correctly described.

\section{Inclusion of a trapping harmonic potential}

We consider now the effect of confinement due to an external anisotropic harmonic potential
\begin{equation} 
 U({\bf r}) = {m\over 2} \left( \omega_{\bot}(x^2+y^2) 
            + \omega_z z^2 \right) \; 
\end{equation}
where $\omega_{\bot}$ is the cylindric radial frequency and $\omega_z$ is the cylindric 
longitudinal frequency. 
The collective dynamics of the system can be described efficiently by the hydrodynamic 
equations, modified by the inclusion of the external potential $U({\bf r})$ 
\cite{book1,book2}, namely 
\begin{eqnarray}
  {\partial n\over \partial t} + 
  {\boldsymbol \nabla} \cdot \left( n \, {\bf v} \right) 
  = 0 \; , 
  \label{dyn1n}
 \\
  m {\partial {\bf v}\over \partial t} + {\boldsymbol \nabla} 
  \left[{1\over 2} m v^2 + \mu[n,a] + U({\bf r})\right] = {\bf 0} \; .   
  \label{dyn2n}
\end{eqnarray}
It has been shown in Ref.~\cite{cozzini} that by assuming a power-law dependence 
$\mu = \mu_0\, n^{\gamma}$ for the chemical potential (polytropic equation of state) from 
Eqs. (\ref{dyn1}) and (\ref{dyn2}) one finds analytic expressions for the collective 
frequencies. 
In particular, for very elongated cigar--shaped traps ($\omega_{\rho}/\omega_z \gg 1$) the
collective radial breathing mode frequency $\Omega_{\rho}$ is given by 
\begin{equation}
 \Omega_{\rho} = \sqrt{2(\gamma +1)} \, \omega_{\rho} \; , 
\end{equation}
while the collective longitudinal breathing mode $\Omega_z$ is 
\begin{equation} 
 \Omega_{z} = \sqrt{3\gamma +2\over \gamma +1} \, \omega_{z} \; .  
\end{equation}
In our problem we introduce an effective polytropic index $\gamma$ as the logarithmic 
derivative of the chemical potential $\mu$, that is
\begin{equation}
 \gamma = {n\over \mu} {\partial \mu\over \partial n} = 
          {2xf_g'(x)+x^2f_g''(x)\over f_g(x)+xf_g'(x)} \; ,  
 \label{poly}
\end{equation}
where $f_g(x)$ is given by Eq. (\ref{liq_fit}). 
This approach has been very successful \cite{nick} in the study the the 
experimentally-observed \cite{grimm} breathing modes of a two-component Fermi gas of 
$^6$Li atoms in the BCS-BEC crossover. 
Indeed in Ref. \cite{nick} we have suggested relevant deviations to the mean-field results, 
which have been subsequently confirmed by improved experiments \cite{grimm2}. 

\begin{figure}
 \epsfig{file=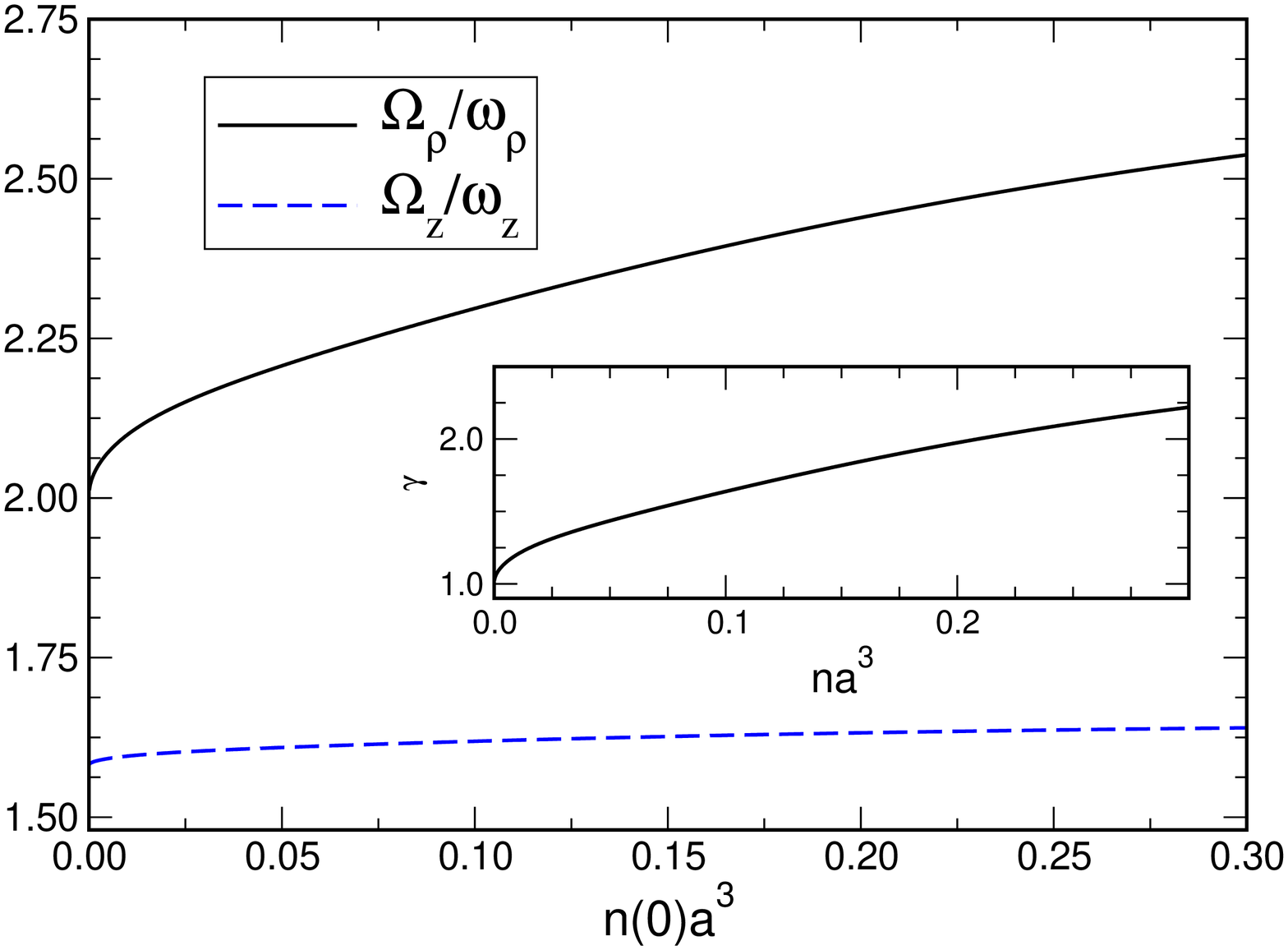,width=8.cm,clip=}
 \caption{(Color online) Breathing mode frequencies of the Bose gas of HS under 
          strong anisotropic axially-symmetric harmonic confinement. 
          $\Omega_{\rho}$ is the frequency of the radial breathing mode and 
          $\Omega_z$ is the frequency of the axial breathing mode. 
          Here $n(0)a^3$ is the local gas parameter with $n(0)$ the gas density at 
          the center of the trap.
          Inset: effective polytropic index $\gamma$ as a function of the gas
          parameter.}
 \label{breathing}
\end{figure}

In Fig. \ref{breathing} we report the frequencies $\Omega_{\rho}$ and $\Omega_z$ of 
breathing modes as a function of the gas parameter $n(0)a^3$, where $n(0)$ is the 
density at the center of the strongly-anisotropic harmonic trap. 
The figure shows a relevant change in the scaled radial frequency 
$\Omega_{\rho}/\omega_{\rho}$ that is a direct consequence of the fact that the 
effective polytropic index $\gamma$ increases from $\gamma \simeq 1$ in the weak-coupling 
regime to $\gamma\simeq 2.2$ in the strong-coupling regime as shown in the inset of 
Fig.~\ref{breathing}. 

\section{Conclusions}

The properties of bulk systems of HS for a wide range of the gas parameter $na^3$, going 
from the dilute gas to the solid phase, have been investigated with the exact $T=0$ PIGS
Monte Carlo methods. 
Our results for the energy per particle turn out to be in good agreement with previous 
calculations, performed with different Monte Carlo techniques, in the gas parameter range 
in which they were aviable \cite{kalo,boro2}.
We have found that recent beyond mean--field approximations are compatible with our Monte
Carlo data up to $na^3\simeq 10^{-3}$. 

We have then fitted our PIGS data via polynomials functions, Eqs. (\ref{liq_fit}) and 
(\ref{liq_fit0}), which have been then used to locate the gas-liquid transition with a 
standard Maxwell construction.
Our analytical fit extends the range of applicability of previous equation of state 
\cite{boro1} up to the freezing point, and beyond it in the metastable region. 
We have computed also the condensate fraction $n_0/n$ in the whole considered gas 
parameter range. 
In particular, we have found that the condensate fraction is zero in the solid phase, in 
agreement with what happens in the solid phase of systems interacting with more realistic 
potentials such as $^4$He \cite{worm,vita}. 
We have provided an analytical fit also for $n_0/n$ showing that, as in Ref.~\cite{boro2}, 
the Bogoliubov approximation overestimates the condensate fraction for $na^3$ larger than 
$10^{-3}$, while the recent improved perturbative approach of Ref.~\cite{yuka} extends 
the predictive region of mean field approaches of about an order of magnitude, up to 
$na^3\simeq10^{-2}$.

The fit of PIGS data are useful in order to derive other relevant properties of the bulk 
system, such as the chemical potential and the pressure.
By means of the zero-temperature hydrodynamics equations of superfluids it is indeed 
possible to obtain other relevant physical quantities. 
In particular we have calculated the sound velocity for gas parameters up to $0.3$. 
This is relevant also because PIGS cannot give direct access to dynamical properties of 
the system. 
Some qualitative informations about the excitation spectrum can be recovered via the 
Feynman's approximation: the low-wave vector limit of such approximate spectrum agrees with
the linear phononic dispersion obtained from the hydrodynamic equation of superfluids 
(\ref{dyn1}) and (\ref{dyn2}) with the equation of state (\ref{liq_fit}).
More quantitative results on the excitation spectrum, can be obtained by computing the 
intermediate scattering functions via PIGS and then by analytically continuing them with 
inversion methods, like GIFT \cite{gift} for example, in order to recover the dynamical 
structure factor. 
These procedures are typically laborious and require a large amount of computations, and 
they go beyond the aim of this paper; anyway, while writing this paper, we went aware that
a similar study is under investigation \cite{rota}.
It is worthy to note, however, that even in an approximate fashion, also an approximation 
as simple as the Feynman's one is able to capture the emerging of the phonon--roton spectrum
deduced by Landau \cite{land} with the increasing gas parameter. 

Finally, we have shown that analytical expressions of the 
exact equation of state can be useful also for predictions in 
confined systems. The hydrodyanamic equations can be used to calculate 
density profiles and collective modes in various trap configurations \cite{book1,book2}. 
Here we have derived the frequencies 
of the collective breathing modes of an HS gas confined in a 
strongly--anisotropic harmonic trap as a function of the local gas parameter. 
By including a gradient correction in the hydrodynamic 
equations one can rewrite them as a nonlinear 
Schr\"odinger equation (generalized Gross-Pitaevskii equation) \cite{sala-snlse,sala-adh} 
and study other foundamental properties 
like quantized vortices \cite{sala-adh}, solitons \cite{sala-soli} and 
shock waves \cite{sala-shock}. 

\section*{Acknowledgments}

The authors thank F. Ancilotto, D.E. Galli, R. Rota, and F. Toigo for useful discussions. 
The authors acknowledge for partial support Universit\`a di Padova (Research Project 
"Quantum Information with Ultracold Atoms in Optical Lattices"), Cariparo Foundation 
(Excellence Project "Macroscopic Quantum Properties of Ultracold Atoms under Optical 
Confinement"), and Ministero Istruzione Universit\`a Ricerca (PRIN Project "Collective 
Quantum Phenomena: from Strongly-Correlated Systems to Quantum Simulators").

\end{document}